\documentstyle [12pt,aaspp4]{article}
\def\beq{\begin{equation}}
\def\eeq{\end{equation}}
\def\bey{\begin{eqnarray}}
\def\eey{\end{eqnarray}}
\def\pppm{\rm P^3M}
\def\mpc{\,h^{-1}{\rm {Mpc}}}

\def\br{{\bf r}}
\def\bk{{\bf k}}

\def\gs{\mathrel{\raise1.16pt\hbox{$>$}\kern-7.0pt
\lower3.06pt\hbox{{$\scriptstyle \sim$}}}}
\def\ls{\mathrel{\raise1.16pt\hbox{$<$}\kern-7.0pt
\lower3.06pt\hbox{{$\scriptstyle \sim$}}}}
\def\gtsima{$\; \buildrel > \over \sim \;$}
\def\ltsima{$\; \buildrel < \over \sim \;$}
\def\prosima{$\; \buildrel \propto \over \sim \;$}
\def\gsim{\lower.5ex\hbox{\gtsima}}
\def\lsim{\lower.5ex\hbox{\ltsima}}
\def\simgt{\lower.5ex\hbox{\gtsima}}
\def\simlt{\lower.5ex\hbox{\ltsima}}
\def\simpr{\lower.5ex\hbox{\prosima}}

\begin{document}
\title {
Accurate determination of the Lagrangian bias for the dark matter
halos}
\author {Y.P. Jing} 
\affil{Research Center for the Early Universe,
School of Science, University of Tokyo, Bunkyo-ku, Tokyo 113-0033, Japan}
\affil {e-mail: jing@utaphp2.phys.s.u-tokyo.ac.jp}
\received{---------------}
\accepted{---------------}

\begin{abstract}
  We use a new method, the cross-power spectrum between the linear
  density field and the halo number density field, to measure the
  Lagrangian bias for dark matter halos. The method has several
  important advantages over the conventional correlation function
  analysis. By applying this method to a set of high-resolution
  simulations of $256^3$ particles, we have accurately determined the
  Lagrangian bias, over 4 magnitudes in halo mass, for four scale-free
  models with the index $n=-0.5$, $-1.0$, $-1.5$ and $-2.0$ and three
  typical CDM models. Our result for massive halos with $M\ge M_\ast$
  ($M_\ast$ is a characteristic non-linear mass) is in very good
  agreement with the analytical formula of Mo \& White for the
  Lagrangian bias, but the analytical formula significantly
  underestimates the Lagrangian clustering for the less massive halos
  $M<M_\ast$. Our simulation result however can be satisfactorily
  described, with an accuracy better than 15\%, by the fitting formula
  of Jing for Eulerian bias under the assumption that the Lagrangian
  clustering and the Eulerian clustering are related with a linear
  mapping.  It implies that it is the failure of the Press-Schechter
  theories for describing the formation of small halos that leads to
  the inaccuracy of the Mo \& White formula for the Eulerian bias. The
  non-linear effect in the mapping between the Lagrangian clustering
  and the Eulerian clustering, which was speculated as another
  possible cause for the inaccuracy of the Mo \& White formula, must
  be negligible compared to the linear mapping. Our result indicates
  that the halo formation model adopted by the Press-Schechter
  theories must be improved.

\end{abstract}


\section {Introduction} 

Galaxies and clusters of galaxies are believed to form within the
potential wells of virialized dark matter (DM) halos. Understanding
the clustering of DM halos can provide important clues to understanding
the large scale structures in the Universe.  A
number of studies have therefore been carried out to obtain the
two-point correlation function $\xi_{hh}$ of DM halos. Two distinctive
approaches are widely adopted.  One is analytical and is based on the
Press-Schechter (PS) theories (e.g. Kashlinsky \cite{k87}, \cite{k91};
Cole \& Kaiser \cite{ck89}; Mann,
Heavens, \& Peacock \cite{mhp93}; Mo \& White \cite{mw96}, hereafter
MW96; Catelan et al. \cite{clmp98}; Porciani et al. \cite{pmlc98}).
The other is numerical and is based on N-body simulations (e.g. White et
al. \cite{wfde87}; Bahcall \& Cen
\cite{bc92}; Jing et al. \cite{jmbf93}; Watanabe, Matsubara, \& Suto
\cite{wms94}; Gelb \& Bertschinger \cite{gb94}; Jing, B\"orner, \&
Valdarnini \cite{jbv95}; MW96; Mo, Jing, \& White \cite{mjw96}; Jing
\cite{j98}; Ma \cite{m98}). The up to date version of the analytical studies is
given by MW96, which states that $\xi_{hh}(r,M)$ of halos with a mass
$M$ is proportional to the DM correlation function $\xi_{mm}(r)$ on
the linear clustering scale ($\xi_{mm}(r)\ll 1$), i.e. $\xi_{hh}(r,M)=
b^2(M)\xi_{mm}(r)$, with the bias factor
\beq\label{mw96b}
 b_{mw}(M) =1+{\nu^2-1\over \delta_c}\,,
\eeq 
where $\delta_c=1.68$, $\nu\equiv \delta_c/\sigma(M)$, and $\sigma(M)$
is the linearly evolved rms density fluctuation of top-hat spheres
containing on average a mass $M$ (see MW96 and references therein for
more details about these quantities). The subscript $mw$ for $b(M)$ in
Eq.(\ref{mw96b}) denotes the result is analytically derived by MW96.
On the other hand, the most accurate simulation result was recently
presented by our recent work (Jing \cite{j98}), where we studied $\xi_{hh}$
for halos in four scale-free models and three CDM models with the help
of a large set of high-resolution N-body simulations of $256^3$
particles. Our result unambiguously showed that while the bias is
linear on the linear clustering scale, the bias factor given by MW96
significantly underestimates the clustering for small halos with
$\nu<1$. Our simulation results both for the CDM models and the 
scale-free models can be accurately fitted by
\beq\label{fitting}
b_{fit}(M)=({0.5\over \nu^4}+1)^{(0.06-0.02n)} (1+{\nu^2-1\over
  \delta_c})
\,,
\eeq
where $n$ is the index of the linear power spectrum $P_m(k)$
at the halo mass $M$
\beq \label{neff}
n ={d \ln P_m(k) \over d \ln k}\Bigg|_{k={2\pi\over R}}; \hskip 1.5cm
R=\Bigl({3M\over 4\pi\bar\rho}\Bigr)^{1/3}\,. 
\eeq 
In the above equation $\bar\rho$ is the mean density of the
universe. 

MW96 derived their formula in two steps. First they obtained the
bias factor $b_{mw}^L(M)$ in the Lagrangian space using the PS
theories. The Lagrangian bias reads,
\beq\label{mw96bl}
 b^L_{mw}(M) ={\nu^2-1\over \delta_c}\,.
\eeq 
But the bias that is observable is in the Eulerian space. MW96
obtained the Eulerian bias (Eq.\ref{mw96b}) with a {\it linear
  mapping} from the Lagrangian clustering pattern,
$b_{mw}(M)=b_{mw}^L(M)+1$ (cf.  Catelan et al. \cite{clmp98}). From
their derivation, we conjectured in Jing (\cite{j98}) that two
possibilities could have failed the MW96 formula for small halos. The
first possibility is that the PS theories are not adequate for
describing the formation of small halos. The halo formation in the PS
theories is uniquely determined by the local peak height through the
spherical collapse model, while in reality, especially the formation
of small halos, can be significantly influenced by the non-local tidal
force (e.g. Katz, Quinn, \& Gelb \cite{kqg93}; Katz et
al. \cite{kqbg94}).  A recent analysis by Sheth \& Lemson
(\cite{sl98}) for simulations of $100^3$ particles also gave some
evidence that the Lagrangian bias of small halos has already deviated
from the MW96 prediction (Eq.\ref{mw96bl}).  Possible invalidity of
the linear mapping, the second possibility that fails the MW96
formula, was recently discussed by Catelan et al. (\cite{cmp98}). They
pointed out that the linear mapping might not be valid for small halos
because of large scale non-linear tidal force. All these evidences are
important but very preliminary and qualitative. It would be important
to find out if one or a combination of the two possibilities can {\it
  quantitatively} explain the failure of the MW96 formula.

In this Letter, we report our new determination for the Lagrangian
bias factor $b^L(M)$ using the simulations of Jing (\cite{j98}).  We
use a novel method, which we call the cross-power spectrum between the
linear density field and the halo number density field, to measure the
bias factor. The method has several important advantages over the
conventional correlation function (CF) estimator. Applying this method
to our high-resolution simulations yields a very accurate
determination of $b^L(M)$ for halo mass over four magnitudes both in
scale-free models and in CDM models. Our result of $b^L(M)$ can be
accurately represented by $b_{fit}(M)-1$, which quantitatively
indicates that it is the failure of the PS theories for describing the
formation of small halos that results in the failure of the MW96
formula.  This result has important implications for the PS theories.

An independent, highly related work by Porciani et al. (\cite{pcl98})
appeared when the present work had been nearly finished. They measured
the Lagrangian bias for two scale-free simulations ($n=-1$ and $n=-2$)
of $128^3$ particles by fitting the halo two-point correlation
function with the linear and second-order terms (corresponding to the
coefficients $b_1^L$ and $b_2^L$ in Eq.~\ref{dhdm}).  They concluded
that the failure of the MW96 formula essentially exists in the
Lagrangian space and that their result can be reasonably described by
$b_{fit}(M)-1$. While our present study ($n=-1$ and $n=-2$) confirms
their result in these aspects, our simulations have significantly
higher resolution (a factor 8 in mass) that is essential for a robust
accurate measurement of clustering for small halos. Moreover, we
explore a much larger model space and use a superior measurement
method. In addition, there is some quantitative difference between
their measured bias and ours which will be discussed in Sect.~3.

We will describe and discuss the cross-power spectrum method in
Sect.~2, where we will also briefly describe our halo catalogs.  Our
measurement results will be presented and compared, in Sect.~3, to the
analytical formula Eq.(\ref{mw96bl}) and the fitting formula
Eq.(\ref{fitting}) for the Eulerian bias. In Sect.~4, we will
summarize our results and discuss their implications for the PS
theories and for galaxy formation.

\section{Methods and simulation samples}
We use the fluctuation field $\delta (\br)=\rho(\br)/\overline \rho
-1$ to denote the density field $\rho(\br)$, where $\overline \rho$ is
the mean density. Smoothing both the halo number density field and the
linear density field over scales much larger than the halo Lagrangian
size, we assume that the smoothed halo field of $\delta_h(\br)$ can be
expanded in the smoothed linear density field $\delta_m(\br)$ (Mo et
al. \cite{mjw97}),
\beq \label{dhdm}
\delta_h(\br) = b_1^L \delta_m(\br) + {1\over 2} b_2^L 
\delta^2_m(\br) + ...\,.
\eeq 
This general assumption, especially to the first order, has been
verified by previous simulation analysis (e.g. MW96; Mo, Jing, \&
White \cite{mjw97}; Jing \cite{j98}). This expansion is however
naturally expected in the PS theories, with the first coefficient
$b_1^L\equiv b^L$ given by Eq.(\ref{mw96bl}) and the higher order
coefficients by MW96, Mo et al. (\cite{mjw97}), and Catelan et
al. (\cite{clmp98}).

By Fourier transforming Eq.(\ref{dhdm}), multiplying the both sides by
$\delta^{\ast}_m (\bk)$, and taking an ensemble average, we get the
cross-power spectrum $P_c(\bk)\equiv \langle \delta_h (\bk)
\delta^{\ast}_m (\bk)\rangle$.  Because the bispectrum of the linear
density field vanishes for Gaussian fluctuations, the second term at the
right hand side is zero. Therefore we have
\beq\label{crops} 
P_c(k) =b^L P_m(k)+ {\rm (3rd~ and ~higher ~ order ~ terms)}\,,  
\eeq
where $P_m(k)$ is the linear power spectrum.  This equation serves
as a base for our measuring $b^L$.  The linear density field
$\delta_m(\bk)$ is known when we set the initial condition for the
simulations. The halo density field $\delta_h (\bk)$ can be easily
measured for a sample of the DM halos with the FFT method.  The
ensemble average in Eq.(\ref{crops}) can be replaced in measurement by
the average over different modes within a fixed range of the
wavenumber $k$. Thus both $P_c(k)$ and $P_m(k)$ can be easily
measured. The bias factor $b^L$ is just the ratio of these two
quantities, if higher order corrections are small and can be
neglected.

This method has several important advantages over the conventional CF
analysis. On the linear scale where the clustering of small halos is
weak (about 10\% of the mass correlation), the cross-power spectrum
can be estimated accurately, because it does not suffer from the
finite volume effect or the uncertainty of the mean halo number
density.  The errors of $P_c(k)$ and $P_m(k)$ are uncorrelated among
different $k$-bins for a Gaussian field, which ease our error estimate
for $b^L$.  The second order correction ($b_2^L$-term) vanishes in the
cross-power spectrum. The method yields a determination of $b_1^L$,
not the square of $b^L$, thus we can see if the bias factor $b_1^L$ is
positive for $M/M_\ast>1$ and negative for $M/M_\ast<1$ as
Eq.(\ref{mw96bl}) predicts, where $M_\ast$ is the characteristic mass
defined by $\sigma(M_\ast)=\delta_c$. All these attractive features do not
present in the CF analysis. An additional interesting feature is that
the shot noise of the finite number of halos is greatly suppressed in
the estimate of $P_c(k)$, because the linear density field dose not
contain any shot noise and the cross-power spectrum of this field
with the shot noise (of a random sample) is zero in the mean. It is
quite different from the (self) power spectrum estimate of the halos
which we must correct for the shot noise $1/N$ ($N$ is the number of
halos). In the next section, we will quantitatively show the shot
noise effect is indeed negligible in our measurement of $b^L$ even for
a sample containing as many as $\sim 100$ halos.

The halo catalogs analyzed here are the same as those used in Jing
(\cite{j98}). The cosmological models, the simulations, and the halo catalogs
were described in detail by Jing (\cite{j98}). Here we only briefly
summarize the features that are relevant to the present work.  The
catalogs were selected, with the friends-of-friends algorithm with the
linking length 0.2 times the mean particle separation, from a set of
$\pppm$ N-body simulations of $256^3$ particles.  The simulations
cover four scale-free models $P_m(k)\propto k^n$ with $n=-0.5$, $-1.0$,
$-1.5$ and $-2.0$ and three typical cold dark matter (CDM) models
which are SCDM, LCDM and OCDM models respectively.  Each of the $n\ge
-1.5$ scale-free simulations has seven outputs, and that of the $n=-2$
has eight outputs. The CDM models are simulated with two different box
sizes, $100$ and $300\mpc$.  Three to four realizations were run for
each model and for each box size in the case of the CDM models. In
order to study the halo distribution in the Lagrangian space, we
trace back each of the halo members to its initial position before the
Zel'dovich displacement, i.e. the position in the Lagrangian
space. The position of a halo is defined as
the center of the mass of its members in the Lagrangian space. In this
way our halo catalogs in the Lagrangian space are compiled.

\section{The Lagrangian bias parameter}
We present our results for the linear clustering scales where the
variance $\Delta^2(k)\equiv k^3P_m(k)/2\pi^2$ is less than
$\Delta^2_{max}$. In this paper, we take $\Delta^2_{max}=0.5$, but our
results change little if we take $\Delta^2_{max}=0.25$ or
$\Delta^2_{max}=1.0$.  Figure 1 shows the ratio of the cross-power
spectrum $P_c(k)$ to the linear power spectrum $P_m(k)$ as a function $k$
for two halo masses in the $n=-0.5$ scale-free model. The ratios, i.e.
the bias factors, do not depend on the scale $k$, so the linear bias
approximation is valid.  The bias factor is positive for the large
halos of $M=13 M_\ast$, but negative for the small halos of
$M=0.16 M_\ast$, consistent with the MW96 prediction Eq.(\ref{mw96bl}).
Here we only show two examples, but the above features are
found in all the models. 

To quantitatively examine the shot noise effect caused by the finite
number of the halos (\S 2), we repeat our calculation for random
samples. For each halo sample, we generate ten random samples. The
mean bias factor calculated for the random samples is zero, as
expected.  More interesting is that the fluctuation of the bias factor
between the random samples is always small compared to the halo bias
factor. This is shown in Figure 1, where the dashed lines are the
$1\sigma$ upper limits of the shot noise.  The catalog of $M=13M_\ast$
contains only $\sim 200$ halos in each realization. Even in this case,
the shot noise leads to an uncertainty of only $\sim 10\%$ of the
halo bias at every $k-$bin (except for the first bin). The shot
noise is indeed effectively suppressed in our cross-power spectrum
measurement.

From Figure 1, we know that the Lagrangian bias $b^L$, on the linear
clustering scale, is a function of the halo mass $M$ only. The
self-similar nature of the scale-free models makes the bias depend
only on $M/M_\ast$.  Three panels of Figure 2 show our result of
$b^L_1 (M/M_\ast)$ for the models $n=-0.5$, $-1.0$ and $-2.0$ at
different output times.  We do not plot, for the limited space, our
result for the $n=-1.5$ model, but all of our following discussion
still includes this model.  The excellent scaling exhibited by
$b^L(M/M_\ast)$ at different outputs supports that our measurement is
not contaminated by any numerical artifacts. The bias factor is
negative for $M<M_\ast$, zero for $M=M_\ast$, and positive for
$M>M_\ast$, in agreement with the MW96 formula. More quantitatively,
the MW96 formula describes very well the Lagrangian clustering for
large halos $M\gs M_\ast$, but systematically underestimates (more
negative bias) for small halos $M\ls M_\ast$. If the linear mapping is
valid, i.e. $b=b^L+1$, these results are fully consistent with our
previous finding (Jing \cite{j98}) that the MW96 formula
systematically underestimates the Eulerian bias for small halos. More
interestingly, with the linear mapping assumption and with our fitting
formula Eq.(\ref{fitting}) for the Eulerian bias, we can get a
Lagrangian bias $b^L_{fit}\equiv b_{fit}-1$ that agrees very well with
our simulation results (the solid lines on Fig.~2). So the inaccuracy
of the MW96 Eulerian bias formula Eq.~(\ref{mw96b}) already exists in
their derivation for the Lagrangian bias.  Our results for the $n=-1$
and $n=-2$ models are qualitatively in good agreement with Porciani et
al. (\cite{pcl98}). Quantitatively, we note that their Lagrangian bias
is significantly lower than the MW96 formula for $M>M_\ast$ in the
$n=-1$ model, in contrast to the good agreement we find with the MW96
formula for all models for $M> M_\ast$.

Our result for the LCDM model is shown in the lower-right panel of
Figure 2. Comparing to the MW96 analytical formula Eq(\ref{mw96b})
and the fitting formula Eq(\ref{fitting}), we have the same conclusion
as we got for the scale-free models: the analytical formula
underestimates the Lagrangian bias value for small halos and the
fitting formula agrees quite well with the simulation results after
the difference between the Eulerian and the Lagrangian spaces is
considered with the linear mapping. The SCDM and OCDM models give
essentially the same results, and we omit their plots.

Our fitting formula for the Eulerian bias can accurately
describe the Lagrangian bias under the assumption of the linear
mapping. The difference between the fitting formula and the simulation
result is generally less than $\sim 15\%$ or $2\sigma$ ($\sigma$ is
the error derived from the different realizations) except for one bin
of the smallest halos in the $n=-0.5$ model.  The simulation result is
slightly higher in the $n=-0.5$ model but lower in the $n=-2.0$ model
(both about 15\%) than the fitting formula for small halos ($M\ll
M_\ast$).  The difference could come from the possible effect of
non-linear mapping (Catelen et al. \cite{cmp98}) and/or from
higher order contributions which depend on measurement methods (for
the Eulerian bias we used the CF analysis). We will address this
problem more closely in a future paper.

\section{Discussion and conclusions}
In this Letter, we use a new method, the cross-power spectrum between
the linear density field and the halo number density field, to
determine the Lagrangian bias. The method has several apparent
advantages over the conventional correlation function estimator in
determining the bias factor. Applying this method to the halo catalogs
of Jing (\cite{j98}), we find that the Lagrangian bias is linear on the
linear clustering scale. The Lagrangian bias $b^L(M)$ is
positive for halo mass $M>M_\ast$, zero for $M=M_\ast$, and negative
for $M<M_\ast$, qualitatively consistent with the MW96 prediction for
the Lagrangian bias. Quantitatively, our simulation results of $b^L(M)$
are in good agreement with the MW96 formula for large halos
$M/M_\ast\gs 1$, but the MW96 formula significantly underestimates
the Lagrangian clustering for small halos $M/M_\ast< 1$. Our
measured Lagrangian bias can be described very well by our fitting
formula for the Eulerian bias Eq.(\ref{fitting}) under the linear
mapping assumption.  Our results therefore unambiguously demonstrate
that the inaccuracy of the MW96 formula for the Eulerian bias already
exists in their derivation for the Lagrangian bias.

A very subtle point is that there exists a small difference ($\ls
15\%$) between our measured Lagrangian bias and our fitting formula
[Eq.(\ref{fitting})] for Eulerian bias after the linear mapping is
applied. This difference could be due to the difference of the
measurement methods or/and the non-linear mapping effect. Our result
however assures that the non-linear effect in the mapping, if any,
must be small compared to the linear mapping. 

The result of this paper has important implications for the
Press-Schechter theories. The spherical collapse model that connects
the halo formation with the density peaks has to be replaced with a
model that can better describe the formation of small halos. This
might be related to solving the long standing problem that the halo
number density predicted by the Press-Schechter theories has a factor
of a few difference from simulation results (e.g., Gelb \&
Bertschinger \cite{gb94}; Lacey \& Cole \cite{lc94}; Ma \&
Bertschinger \cite{mb94}; Jing 1998, unpublished; Tormen \cite{t98};
Kauffmann et al. \cite{ketal98}; Somerville et al. \cite{setal98};
Governato et al. \cite{getal98}; Lee \& Shandarian \cite{ls98}; Jing,
Kitayama, \& Suto \cite{jks99}). Because galaxies are believed to form
within small halos ($M<M_\ast$), solving these problems is of
fundamental significance to studying galaxy formation.

\acknowledgments 

It is also my pleasure to thank an anonymous referee for a
constructive report and the JSPS foundation for a postdoctoral
fellowship.  The simulations were carried out on VPP/16R and VX/4R at
the Astronomical Data Analysis Center of the National Astronomical
Observatory, Japan.

\begin{figure} 
\epsscale{1.0} \plotone{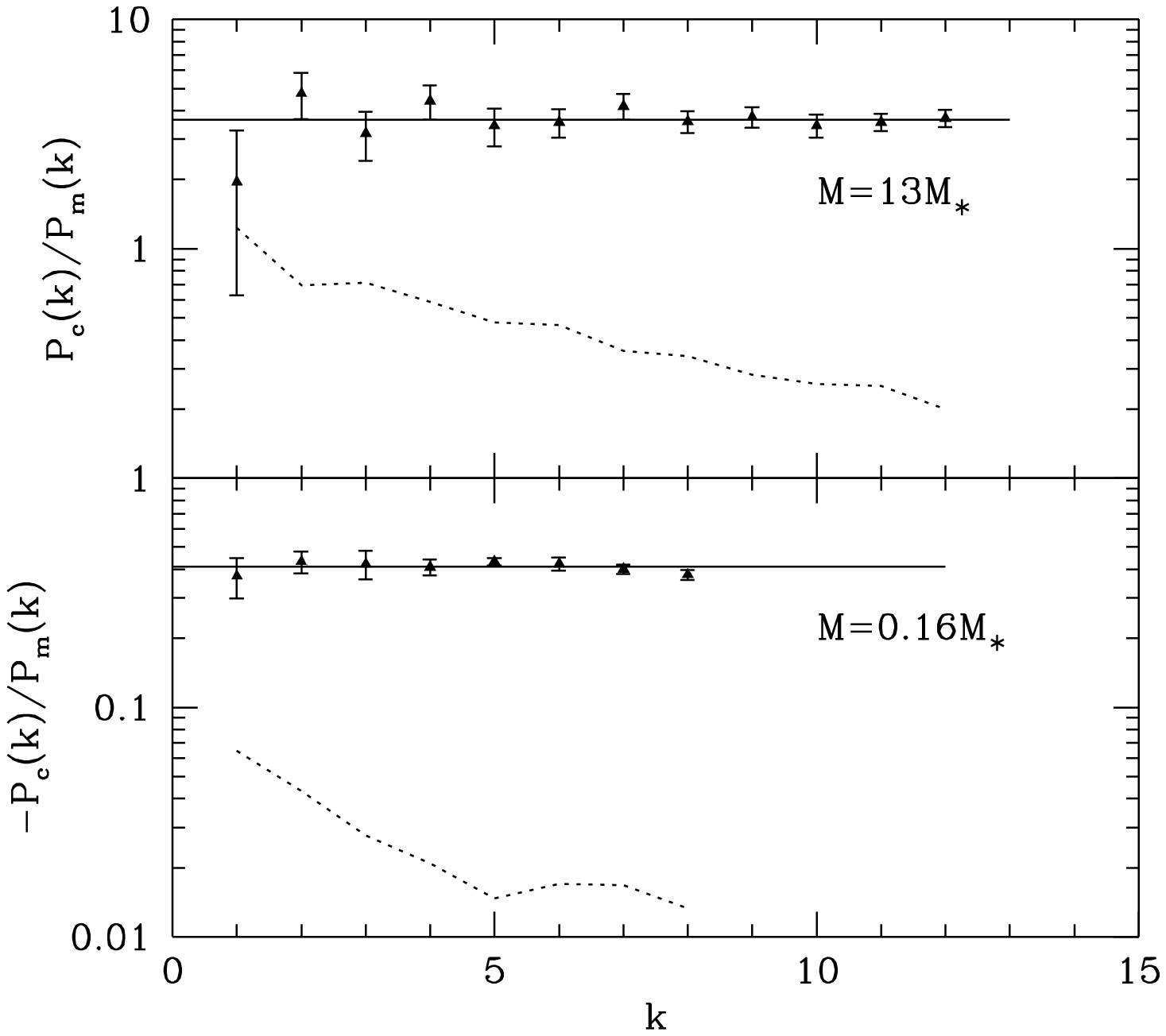} 
\caption{ The ratio of the cross-power spectrum $P_c(k)$ to the linear
density power spectrum $P_m(k)$ in the scale-free model of $n=-0.5$.
The wavenumber $k$ is in units of the fundamental wavenumber of the
simulation. Only the linear clustering regime is considered here. The
upper panel is for halo mass $M/M_\ast=13$, and the lower one is for
$M/M_\ast=0.16$.  In the lower panel, since the ratio $P_c(k)/P_m(k)$
is negative, we plot $-P_c(k)/P_m(k)$. The solid lines are the mean
ratio averaged for different scales. The dashed lines are the
$1\sigma$ upper limits of the shot noise effect caused by the finite
number of halos.}
\label{fig1}\end{figure}

\begin{figure}
\epsscale{0.8} \plotone{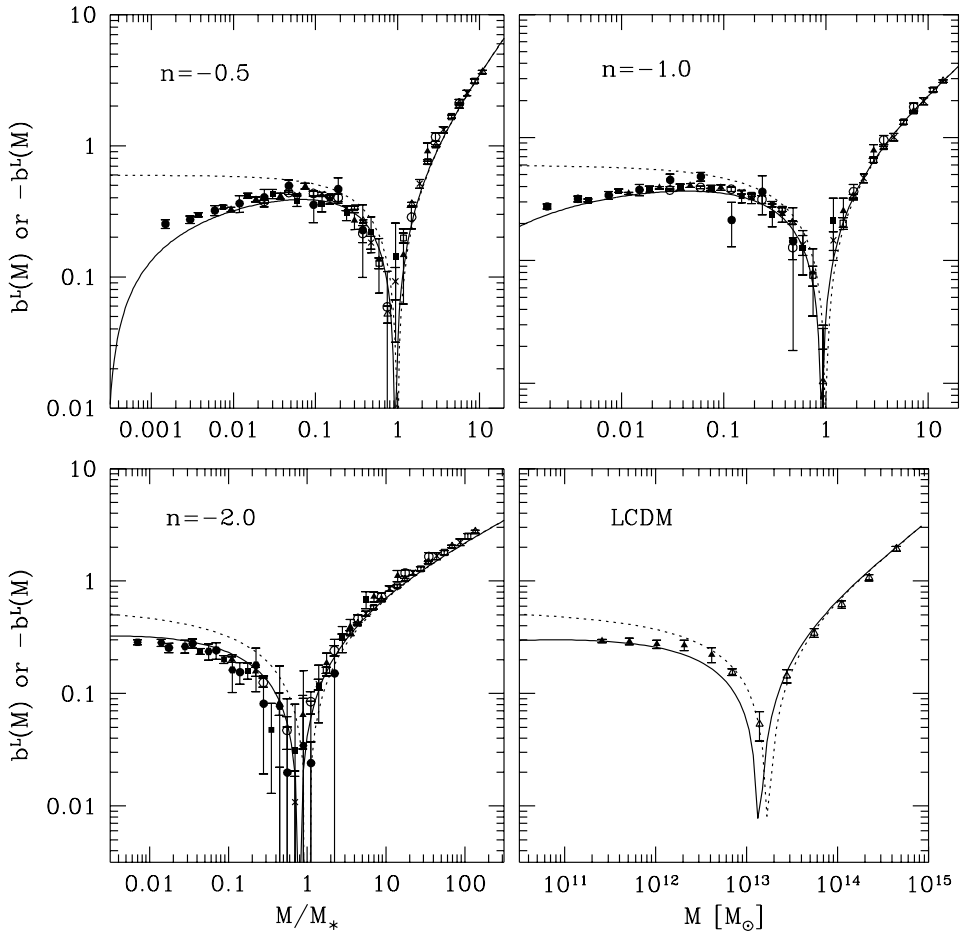}

\caption{ The Lagrangian bias parameter $b^L$ as a function of the
  halo mass $M$. The mass is in units of $M_\ast$ in the scale-free models
  and $M_\odot$ in the LCDM model.  Because $b^L$ is negative for
  $M/M_\ast<1$, we plot $b^L$ for $M/M_\ast>1$ but $-b^L$ for
  $M/M_\ast<1$.  For the scale-free models, the results at seven
  different evolutionary stages are plotted with different
  symbols. From the early to the late outputs, the symbols are
  respectively open triangles, open squares, crosses, open circles,
  solid triangles, solid squares, and solid circles. For $n=-2$, the
  result for a further output (more clustered) at the 1362th time step
  is added with hexagons. It is interesting to note that the results
  from different outputs agree remarkably well. For the LCDM model,
  the result is shown with open triangles for the simulation of
  box size $300\mpc$ and with filled triangles for that of $100\mpc$.
  The dotted lines are the prediction of Mo \& White (\cite{mw96}) for
  the Lagrangian bias, which is in good agreement with the simulation
  results for $M/M_{\ast}\gs 1$ while it significantly underpredicts
  (more negative) for $M/M_{\ast}\ll 1$.  The solid lines are given by our
  fitting formula (Eq.~\ref{fitting}) for the Eulerian bias under the
  linear mapping assumption, which can describe the simulation results
  remarkably well. }\label{fig2}\end{figure}

\end{document}